\documentclass[%
 reprint,
unsortedaddress,
nofootinbib,
 amsmath,amssymb,
 aps,
floatfix,
]{revtex4-2}

\usepackage{comment}
\usepackage{graphicx}
\usepackage{dcolumn}
\usepackage{bm}
\usepackage[dvipsnames]{xcolor}

\begin{document}

\preprint{APS/123-QED}

\title{Minimal cyclic behavior in sheared amorphous solids
}

\author{Chloe W. Lindeman}
\email{cwlindeman@uchicago.edu}
\author{Sidney R. Nagel}%
\affiliation{%
 Department of Physics and The James Franck and Enrico Fermi Institutes \\ 
 The University  of  Chicago, 
 Chicago,  IL  60637,  USA.\\
\\
}%

\date{\today}
 
\begin{abstract}
Although jammed packings of soft spheres exist in potential energy landscapes with a vast number of minima, when subjected to cyclic shear they may revisit the same configurations repeatedly. Simple hysteretic spin models, in which particle rearrangements are represented by interacting spin flips called hysterons, capture many features of this periodic behavior. Yet it has been unclear to what extent individual rearrangements can be described by such binary objects and how such objects interact with one another. Using a particularly sensitive algorithm, we identify rearrangements in simulated jammed packings and select pairs of rearrangements that undo one another to create periodic cyclic behavior.  We find that the rearrangement pairs surprisingly persist down to the smallest increments in strain, even in the smallest systems we can study. We explore the statistics of these rearrangement pairs and find that there is a relation between the amount of hysteresis and the energy drop and mean-square displacement of the particles; these results are inconsistent with the scaling found in models that treat rearrangements as localized buckling events.  Finally, our analysis shows that there is no clean distinction between the ``core'' of an individual rearrangement and the interactions between rearrangements. These results offer insight into how complex systems such as amorphous solids can reach a limit cycle.
\end{abstract}

\maketitle  

\section*{Introduction}

When strained past the elastic limit, jammed packings of soft spheres exhibit plastic particle rearrangements: instabilities that allow the system to fall irreversibly into a new configuration in the energy landscape~\cite{falk1998dynamics}. Despite the plastic nature of these instabilities, a wide range of experiments and simulations have shown that cyclic deformation may lead to a repeating sequence of rearrangements~\cite{keim2013yielding, regev2013onset, fiocco2013oscillatory, perchikov2014variable, keim2014mechanical, fiocco2014encoding, royer2015precisely, lavrentovich2017period}. Previous studies have measured a variety of physical characteristics of the rearrangement events including their spatial structure 
and the vanishing of the shear modulus and lowest-frequency mode upon approaching an instability~\cite{maloney2004subextensive, maloney2004universal, maloney2006amorphous, shimada2018spatial, xu2017instabilities, ruan2022predicting, xu2023instabilities}. However, these quantities provide little insight into the emergence of the periodic limit cycles.

\begin{figure}
\includegraphics[width=8.6cm]{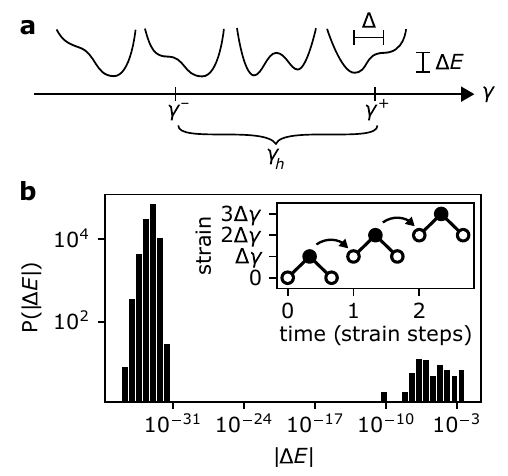}
\caption{(a) Hysteresis produced by a double-energy well modified by applied strain. At two critical values of strain, $\gamma^-$ and $\gamma^+$, the left and right wells flatten out, respectively; two stable minima exist for strains between these values. The distance $\gamma_h \equiv \gamma^+ - \gamma^-$ is a measure of hysteresis. Changes in particle position $\Delta$ and energy $\Delta E$ at an instability are depicted schematically. (b) Distribution, $P(|\Delta E|)$, of changes in energy upon returning to the same strain after one strain step for $N=31$, $\phi=0.87$ and $\Delta \gamma=10^{-5}$. Reversible events have $|\Delta E|< 10^{-31}$ and are well separated from true rearrangements which have $|\Delta E|>10^{-12}$. Inset shows method to find $|\Delta E|$ by comparing energies of adjacent open circles. 
}
\label{fig:reversibility}
\end{figure}

By contrast, Falk and Langer considered the possibility that each rearrangement is a transition in a bistable system that can be undone upon reversal of the shear direction~\cite{falk1998dynamics}. As depicted schematically in Fig.~\ref{fig:reversibility}a, this leads to a hysteretic response so that the current state depends on the history of deformation. Models based on hysteretic two-state systems, called $\textit{hysterons}$, were first studied in the context of magnetic materials and Ising models~\cite{preisach1935magnetische, sethna1993hysteresis} and have become widely used to model periodic particle rearrangements in cyclically driven disordered materials~\cite{mungan2019networks, keim2020global, terzi2020state, van2021profusion, szulc2022cooperative, shohat2022memory}.


Though independent hysterons can lead to exceedingly complex trajectories through phase space~\cite{terzi2020state, van2021profusion, bense2021complex, shohat2022memory, mungan2022putting}, they cannot produce the long transients or multi-cycle periodicity observed in jammed systems without the presence of interactions between hysterons~\cite{lavrentovich2017period, keim2021multiperiodic, lindeman2021multiple, szulc2022cooperative}. The notion of such interactions necessitates that one be able to distinguish between a rearrangement --- the ``core'' of a single hysteron --- and the interactions --- the ``dressing'' --- between two rearranging entities. While paired transitions have been measured and analyzed in a meso-scale model system, to our knowledge, there exist no such measurements in amorphous solids~\cite{kumar2022mapping}. 
Analysis of rearrangement pairs is thus necessary to uncover the nature of inter-hysteron interactions so as to understand the connection between amorphous solids and simple models based on hysteretic spins. 

Here we analyze rearrangement pairs in packings of $N$ discs in a two-dimensional box with periodic boundary conditions. We focus on small system sizes, $7 \le N\le 1021$, with the aim of resolving each rearrangement. To isolate the minimal elements for periodic response, we restrict our study to configurations in which the first instability encountered after a packing is created is undone by the first one found upon reversing the shear direction. Such rearrangement pairs are common: we find that about half of all initial rearrangements are undone in this way. The distribution of strain intervals, $\gamma_h$, over which two stable configurations exist, $P(\gamma_h)$, reveals an abundance of pairs with surprisingly small hysteresis. 

For all rearrangements, we measure the correlations between energy drop and total particle displacement; in the case of paired rearrangements, we also correlate these results with $\gamma_h$.  The correlations are incompatible with the scaling found in models based on individual buckling events.   


Finally, to find the core of each hysteron, we measure the number of particles needed to push the system between configurations at strains between the two rearrangements. 
We find no clean distinction between what is the ``core'' of an individual rearrangement event and the ``dressing'' which represents how one rearrangement interacts with another.  This poses problems for analyzing the cyclic response in sheared packings in terms of interacting hysterons. 

\section*{Methods}
We study two-dimensional packings of soft repulsive disks with periodic boundaries under athermal quasistatic shear using the pyCudaPacking package~\cite{morse2014geometric, charbonneau2015jamming}. The energy is defined by:
$$V(r_{ij})=\epsilon(1-r_{ij}/\sigma_{ij})^{\alpha}\Theta(\sigma_{ij}-r_{ij})$$ where $\epsilon$ is the energy scale, $ij$ label particles separated by distance $r_{ij}$, $\sigma_{ij}\equiv \sigma_i + \sigma_j$ is the sum of particle radii, and $\Theta(x)$ is the Heaviside function. Except where specifically mentioned, the exponent $\alpha = 2.5$ (Hertzian contacts) is chosen in order to avoid anomalous instabilities~\cite{xu2023instabilities}.
Packings are created at fixed packing fraction $\phi = 0.95$ and radii are chosen from a log-normal distribution with 20\% polydispersity. 

An initial configuration is generated with random particle positions (infinite temperature) and minimized using the Fast Inertial Relaxation Engine (FIRE). To apply shear, we set the lattice vectors of our periodic boundaries as specified in Appendix A and move each particle affinely before minimizing with FIRE.

In order to identify a rearrangement, we choose a strain step, $\Delta \gamma$, and test for reversibility at every step: we shear forward one strain step then shear backward by the same amount and calculate the magnitude of the energy change, $|\Delta E|$, at the same strain before and after that step, as indicated schematically in the inset to Fig.~\ref{fig:reversibility}b (energy is compared between the connected pairs of open circles). 
After recording $|\Delta E|$, the system is reset to the configuration at the higher value of strain (indicated in the inset by an arrow) and the process is repeated.

As shown in Fig.~\ref{fig:reversibility}b, rearrangements are well-defined; the magnitude of $|\Delta E|$ is either quad-precision error, $< 10^{-31}$, or $>10^{-12}$. For the same set of events, such a separation is not found for other metrics such as the non-affine deformation $D^2_{min}$~\cite{falk1998dynamics} or the energy change between adjacent frames; as shown in Appendix B, in both of those cases, a continuous range of the $x$-axis variable makes it impossible to identify with certainty when a rearrangement has occurred. This precision is essential for identifying unambiguously the very small hysteretic events that characterize the tails of the paired rearrangement distributions.
  
For each packing, we increase the strain until we find a rearrangement. We then decrease the strain until another rearrangement is found and test whether these two rearrangements form a closed orbit. If so, the packing is considered an ``elementary hysteron'' and is saved along with information about both rearrangements. Using a bisection algorithm, the strains of both rearrangements (as well as any unpaired rearrangements) are determined to a resolution of $10^{-7}$ for measurements of rearrangement properties and the rearrangement core. 
If at any point the strain exceeds a threshold $\gamma = 0.3$, or if at any point the entire packing loses rigidity, the packing is discarded. Additionally, if the particle radii are such that a particle interacts with the same neighbor on both sides (an issue only in the smallest packings $N=7$), we choose new radii from the log-normal distribution. 

Our method of finding a rearrangement relies on irreversibility; in particular, a rearrangement pair with $\gamma_h$ smaller than the step size $\Delta \gamma$ would be identified as reversible on the scale of the step size and hence entirely missed in our search protocol. We therefore repeat our search for hysterons using different values of the strain step size: $10^{-5}$ $\le \Delta \gamma \le 10^{-3}$.

\section*{Results and Discussion}

\subsection*{Rearrangement pairs} 

\begin{figure*}[ht]
\centering
\includegraphics[width=17.2cm]{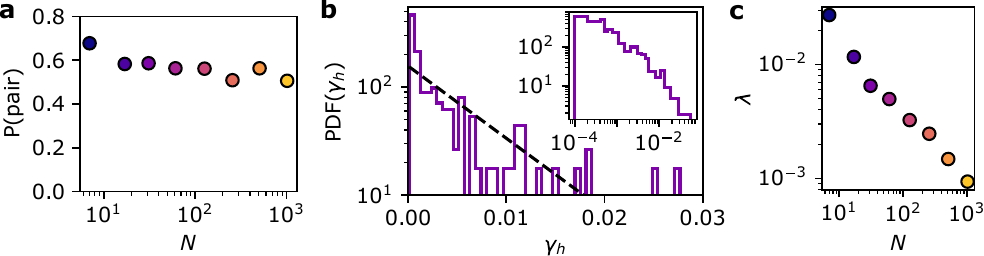}%
\caption{ 
Statistics of paired rearrangements ($\phi =  0.95$ and $\Delta \gamma = 10^{-4}$). (a) Probability $P(\text{pair})$ that a packing immediately falls into a two-rearrangement periodic orbit versus system size $N$. (b) Probability distribution function for 31-particle systems that a rearrangement pair is separated in strain by  hysteresis $\gamma_h$ overlaid by an exponential fit (dashed black line); single-count bins excluded. Inset: same data on log-log axes. 
(c) Characteristic decay strain $\lambda$ of hysteresis $\gamma_h$, obtained from fitting (b) to an exponential, as a function of system size. Colors represent system size and are consistent with Fig.~\ref{figsi:Ncore} in Appendix C. 
}
\label{fig:prob}
\end{figure*}

Using the relative number of packings saved versus those tested, we calculate the probability $P(pair)$ that a randomly generated packing with $N$ particles will result in a rearrangement pair, \textit{i.e.} an elementary hysteron. Figure~\ref{fig:prob}a shows that $P(pair)$ decreases slowly with $N$. 
Our requirement that a rearrangement be immediately undone under strain reversal underestimates the probability that a rearrangement $\textit{could}$ be undone with a more complex path through strain space. The strength of this effect depends on the density (number per strain interval) of total rearrangements, a quantity that increases with system size~\cite{lavrentovich2017period, xu2017instabilities, salerno2013effect}. Hence the probability of finding a rearrangement pair decreases with $N$, as seen in Fig.~\ref{fig:prob}a. 
Moreover, rearrangements which are proportionally larger, whether due to small system size or low pressure~\cite{shimada2018spatial}, are more likely to interact with other instabilities.

For paired rearrangements we compute $P(\gamma_{h})$, the probability that the two rearrangements are separated by a given strain $\gamma_h$. 
As shown in Fig.~\ref{fig:prob}b for $N=31$ and $\Delta \gamma = 10^{-4}$, $P(\gamma_{h})$ is sharply peaked at small $\gamma_h$. Across all systems studied, including two decades of step sizes $\Delta \gamma$ for $N=31$ and two decades of system sizes $N$ for $\Delta \gamma = 10^{-4}$, the peak in $\gamma_h$ was always within noise of the smallest measurable value as set by the step size $\Delta \gamma$. $P(\gamma_{h})$ is approximately exponential: $P(\gamma_{h}) \propto e^{-\gamma_{h}/\lambda}$, providing a characteristic strain scale $\lambda$ over which the probability decays. Figure~\ref{fig:prob}c shows that $\lambda$ decreases with system size $N$; the hysteresis $\gamma_h$ of a rearrangement pair found in a small system is typically larger than that found in a larger system.

The pair-selection algorithm also impacts the measured distribution of hysteresis values. 
The probability of a pair of rearrangements being ``interrupted'' by another rearrangement increases with increasing $\gamma_h$. The chances of such an interruption grows as the number of possible instabilities increases due to increasing $N$ so that $P(\gamma_h)$ is especially tightly peaked around lower values of $\gamma_h$ in larger systems as seen in Fig.~\ref{fig:prob}c. This is consistent with the measurement of a smaller typical strain distance to the first instability when a larger number of particles is probed~\cite{ruan2022predicting} and suggests that other changes to the density of rearrangements, such as via thermal equilibration, would also affect measurements of $\gamma_h$~\cite{sastry1998signatures}. While the density of rearrangements per strain plays a role in the measurement of these quantities, the characteristic \textit{spatial} extent of a rearrangement is also relevant. 

\subsection*{Statistics of rearrangements} 

\begin{figure*}[ht]
\centering
\includegraphics[width=17.2cm]{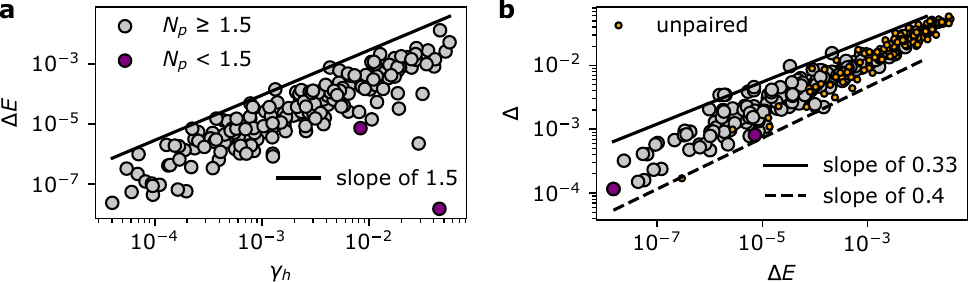}%
\caption{Scaling between different features associated with rearrangements ($N=31$ and $\phi=0.95$). Lines are guides to the eye intended to bound the data; localized rearrangements (that is, with participation number $N_p < 1.5$) are shown in dark purple. (a) Energy drop for hysteron rearrangements measured at $\gamma^+$ versus hysteresis $\gamma_h$. (b) Particle displacements $\Delta$ versus energy drop $\Delta E$ for the same rearrangements (large circles) and for rearrangements which were not undone by the first rearrangement encountered upon changing direction of shear (small orange circles). 
The overlap and similar slope between the paired and unpaired data indicate that the instability associated with one of the pairs in a hysteron is essentially indistinguishable from unpaired instabilities. 
}
\label{fig:hysteresis}
\end{figure*}

Because the hysteresis values $\gamma_h$ range over three decades, we can ask whether coupled rearrangements with small and large $\gamma_{h}$ are fundamentally different from each other. To address this, we compute the energy change and the root-mean-square displacement of all the particles during each rearrangement: $$\Delta \equiv \left( \sum_i \frac{1}{N} ( \Delta r_i)^2\right)^{1/2},$$  where $\Delta r_i$ is the distance the $i^{\textrm{th}}$ particle moved during the instability.
Figures~\ref{fig:hysteresis}a shows that the energy change $\Delta E$ varies smoothly with the hysteresis $\gamma_{h}$. 
Figure~\ref{fig:hysteresis}b shows that root-mean-square displacement $\Delta$ is correlated with $\Delta E$ as well, so that all three quantities are coupled:
paired rearrangements with small $\gamma_{h}$ tend to have both smaller particle motion, $\Delta$, and smaller change in energy, $\Delta E$. The trend between $\Delta$ and $\Delta E$ for unpaired rearrangements is consistent with that for paired rearrangements.



The number of particles participating in each rearrangement, $N_p$, can be measured by calculating the participation ratio of the displacements~\cite{silbert2009normal}. While $N_p$ varies widely from $\sim 1$ to more than half of the system size, it does not, with one exception, clearly correlate with the quantities shown in Fig~\ref{fig:hysteresis}. The exception is rearrangements with $N_p \approx 1$ which behave somewhat differently from other rearrangements as highlighted in Fig~\ref{fig:hysteresis}a; the moving particle in these events tends to correspond to a buckler~\cite{charbonneau2015jamming}. See SI~\cite{SI-ref} for details.

Figure~\ref{fig:hysteresis}b shows an approximate scaling relation between $\Delta E$, the energy drop  associated with a rearrangement, and $\Delta$, the net root-mean-square displacement of particles from the initial 
configuration. We note that while the fluctuations are large, so that it is not unambiguously a power law, the trends are very consistent across a large range of parameters including system size and interaction type (as shown in the SI~\cite{SI-ref}). The exponent, as suggested by the dashed and solid lines, lies between $0.33$ and $0.4$. 

This behavior is reflective of the underlying energy landscape and is not trivial.  One can analytically solve the fully quartic, one-particle spring models that explicitly include both the initial, barely unstable configuration and the global minimum into which the system falls.  These are analyzed in Appendix D and predict power-law exponents that depend on the interaction potential: $1/4$ for harmonic ($\alpha = 2.0$), and $1/5$ for Hertzian interactions ($\alpha = 2.5$).  Both cases are inconsistent with Fig.~\ref{fig:hysteresis}b and give progressively worse agreement as $\alpha$ increases. 

Another approach could be to assume that the energy is described by the so-called fold instability where 
the generic form of the energy precisely at an instability ($\gamma = \gamma^+$) 
is 
\begin{equation}
E = -c \Delta^3,
\label{eq:fold}
\end{equation}
where $c$ is a constant.  This describes accurately the shape of the energy along the lowest lying trajectory out of the initial, barely unstable point, but it does not include the existence of the eventual energy minimum into which the system will come to rest. (It is thus different from the fully quartic models discussed in the preceding paragraph.) One therefore needs to supplement this description with an argument about the location of the new, more stable, minimum to which the system relaxes.

Assuming that the instability is cut off randomly, for example due to newly formed contacts, this cubic behavior should dominate so that the behavior of an ensemble of packings will be $\Delta E \sim \Delta^{3}$ or $\Delta \sim \Delta E^{1/3}$. 
However, while this scaling appears consistent with the data in Fig.~\ref{fig:hysteresis}c, it would be surprising that the fold instability should apply in this regime.  The rational for using it comes from analyzing the behavior only around the previously stable minimum as it becomes unstable~\cite{xu2017instabilities} and should not apply to the regime after the instability where the system has moved to a completely different distant point in phase space. Our analysis of the potential energy as the system falls from the first (barely unstable) well to the other, shown in Appendix E, indicates that a cubic fits only over a very small fraction of the total distance traveled.  

Note that if the fold instability had been applied to either of the single-particle spring models worked out analytically in the Appendix D, it would have failed.  That is, it could have been used to describe accurately the transition \textit{out of} the unstable state, but would have given a very wrong exponent because it does not include the existence of the final state. (We note here that in those spring models, it was important to know what ensemble was being used -- that is, what was being held constant and what was allowed to vary -- in order to obtain the exponents. This is discussed further in the conclusions and in the last section of Appendix D.)  Thus the fold instability without further assumptions does not account for the systematic trends shown in Fig.~\ref{fig:hysteresis}b.


We conclude that the instabilities are collective in accord with the fact that there is a large range of participation ratios that accompany the instabilities. While the trends we present between $\Delta$, $\Delta E$, and $\gamma_h$ are essential to understanding the nature of shear instabilities in disordered materials, they remain unexplained.

\subsection*{Hysteron core} 
We quantify the core of a given rearrangement pair by determining how many particles are needed to switch between the two stable configurations as a function of strain between $\gamma^-$ and $\gamma^+$. At each strain value tested (see Appendix C), we rank the particles by their difference in position between the two configurations, referred to as (–) and (+). Starting in the (–) configuration, we move the particle with the largest difference to its position in the (+) configuration and minimize. We then reset to the (–) configuration and repeat, now moving the \textit{two} particles with the largest differences in positions, then again with four particles, then eight, and so on. (This doubling protocol limits the resolution when $N_{core}$ is large but allows measurements even in large systems.) For each minimization, we test whether we have landed in 
the (+) configuration. The smallest number of particles needed to push the system into the (+) configuration gives us an estimate of the core size $N_\text{core}$ at that value of strain. 

In general, the transition is well behaved: for fewer than $N_\text{core}$ particles moved, the system lands in the (–) basin, while for $N_\text{core}$ or more particles, the system falls into the (+) basin. Occasionally, however, ($\sim 10$\% of packings), the system lands in a third basin for some intermediate number of particles. These packings are excluded from the calculation of average $N_\text{core}$ values.  

\begin{figure*}
\centering
\includegraphics[width=17.2cm]{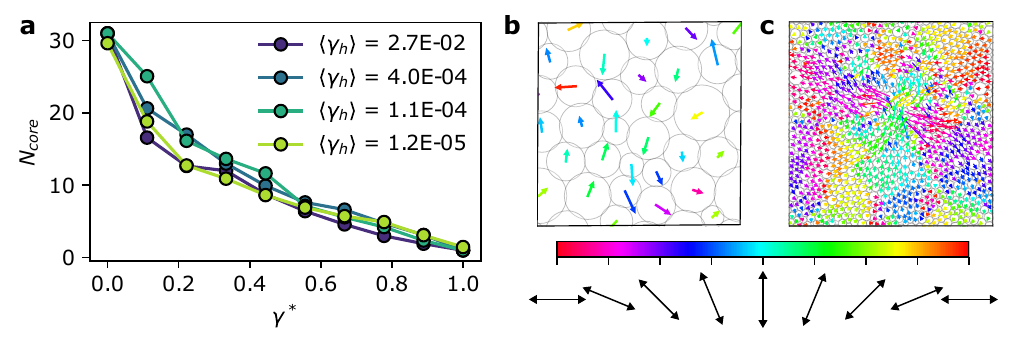}%
\caption{(a)
For systems with $N=31$ at $\phi = 0.95$, the number of particles $N_{\text{core}}$ needed to push the system from one configuration to the other in the double-well potential versus rescaled strain $\gamma^*$ between the two rearrangements.  Different colors show data for curves separated by different amounts of hysteresis $\gamma_h$ as indicated in the legend. (b,c) Particle configurations (circles) and directions of motion during rearrangement (arrows) for systems of size (b) $N=31$ and (c) $N=1021$. Arrow color corresponds to direction particle moves, as shown by the color bar (below). Double-headed arrows represent the fact that positive and negative motion along a particular orientation are represented with same color. Length of arrow is proportional to particle displacement. In (c), for $N=1021$, the colors emphasize the overall quadrulpolar signature of the displacement field. For the smaller system in (b), no clear quadrupolar structure emerges.
}
\label{fig:core}
\end{figure*}

Figure~\ref{fig:core}a shows the number $N_\text{core}$ of particles in the core for different amounts of hysteresis. 
Each curve is averaged over several packings (see Appendix C) and shown as a function of $\gamma^* \equiv (\gamma - \gamma^-)/\gamma_h$. Rearrangement pairs are sorted based on $\gamma_h$ and averages are performed only over hysterons with similar values of $\gamma_h$.

In all curves, the average number of particles needed to be manually moved in order to cause the rearrangement from (–) to (+) varies substantially as $\gamma^*$ is varied. $N_\text{core}$ approaches a substantial fraction of the particles far from the rearrangement and approaches one particle near the rearrangement. The equivalent measurement in the other direction, from (+) to (–), is on average identical when flipped about $\gamma^*=0.5$.

Figures~\ref{fig:core}b and~c show example rearrangements associated with two different system sizes. The larger system size ($N=1021$) shows a quadrupolar displacement field typical of rearrangements in large systems~\cite{maloney2006amorphous}, while this feature is typically not clearly identifiable in smaller systems (for example the $N=31$ case shown in Fig.~\ref{fig:core}b). We note that although some rearrangements may correspond to ``T1 events'' in which two particles lose contact while two others come into contact, such events are not typical; if they were, we might expect to see a feature at $N_{core}=4$ in Fig.~\ref{fig:core}a.  

Because of their origins in the magnetic materials community~\cite{preisach1935magnetische}, hysterons are typically treated like spins. While this binary picture has been useful in attempts to understand general features of cyclically sheared amorphous solids, our results show that it does not hold up under closer scrutiny. In particular, attempts to isolate the particles responsible for switching the system between states as in Fig.~\ref{fig:core}a show that there exists no fixed subset of particles that constitutes the ``hysteron'' at all strains. Instead, the number of particles needed to cause the rearrangement varies systematically with strain. If we consider the local energy landscape as shown schematically in Fig.~\ref{fig:reversibility}a and take $N_\text{core}$ to be a proxy for the energy-barrier height, it is no surprise that this quantity might be large for one strain (that is, when the current configuration is quite stable) and small for another (when the configuration is only marginally stable).



However, this isolated double-well picture breaks down in light of the third energy minimum sometimes accessed when moving particles between the two main configurations. It is an open question whether multi-state hysterons (\textit{i.e.}, hysterons with more than two states) are distinguishable from interacting two-state hysterons~\cite{lindeman2023competition} and, if so, what role such ``hysterons'' play in jammed systems under shear.  The lack of clear separation between the core and dressing of rearrangement pairs, as well as instances of three-well scenarios as described above, suggests that binary spin hysterons are an inadequate model for paired rearrangements in sheared packings. 

\section*{Conclusions}

Modeling cyclic shear as a collection of single reversible plastic events is seemingly at odds with the idea of a jammed packing as a deeply complex glassy system. Yet our results indicate that such double wells not only exist but are in fact relatively easy to find at small values of $\gamma_h$. In the present study, we have not extended the range of shear to analyze how interactions occur when the plastic events occur nearby in space and in shear values so that the hysteresis regions overlap. We have also selected rearrangement events that are capable of being undone, likely excluding more complex rearrangements like those associated with avalanches. In either case, we would expect further complexities. 

Even so, our results may help answer the vexing question of how packings can fall into periodic orbits quickly, even when many rearrangements are involved in a single cycle. One reason can be understood from the data presented above: the statistics of rearrangement pairs restores an effective simplicity.  Interacting pairs tend to scramble one another, delaying the onset of periodicity~\cite{lindeman2021multiple}. Yet to interact, rearrangements must be both within elastic range of each other spatially and also overlapping in strain. Figure~\ref{fig:prob}b shows that most rearrangement pairs have very small $\gamma_h$, making overlap with other rearrangement pairs, and hence interactions, unlikely. With few interactions, the resulting dynamics remain relatively simple, with short transients. 

This argument provides rationale for an emergent periodic response based on local particle rearrangements. 
Another possible approach is to evaluate the volume of all energy basins of an $N$-particle packing. There is a broad distribution of volumes, suggesting that the basins with extremely large size may perhaps play a crucial role in the formation of periodic orbits~\cite{gao2009,hagh2024}. These are two complementary ways of thinking about how a periodic path can be established.  

Though the energy landscape of jammed systems is known to be extraordinarily rugged, for very small systems there may be the expectation that the strain distance between instabilities will be large. Yet a two-dimensional system of $31$ particles corresponds to a fractional particle size which is roughly $31^{-1/2} \sim 0.18$, at odds with the proliferation of $\gamma_h$ values on the order of $10^{-5}$. Moreover, the similarity of $N_{core}$ behavior across different hysteresis amounts, shown in Fig.~\ref{fig:core}, and the consistent trends between $\Delta$, $\Delta E$, and $\gamma_h$ across orders of magnitude, as in Fig.~\ref{fig:hysteresis}, suggest that the features of the double wells are quite unchanged across a wide range of scales. 
What sets these relationships remains an open question.

One important feature of these relationships is the ensemble over which we measure, here quenched disorder generated by starting with different particle configurations and radii; the relationship between particle displacement and energy barrier may depend on this choice of ensemble. A different scaling law was obtained by Ji \textit{et al.} for low-energy excitations in gapped glasses~\cite{ji2022mean}. There, scalings were determined as a function of the minimum frequency $\omega_c$ below which no normal modes exist. Those authors argue that the result holds in ungapped glasses as well. However, that analysis predicts that the energy barrier $E_{loc}$ between minima scales as their separation $X$ to the \textit{sixth} power, $E_{loc} \sim X^6$ or $X \sim E_{loc}^{1/6}$. This suggests that the packing preparation in that work samples from a different ensemble than the preparation reported here.

This discrepancy highlights the importance of specifying what parameters are held fixed and what are allowed to vary during the dynamics of deformation and relaxation shown in Fig.~\ref{fig:hysteresis}; this idea is discussed further in Appendix D. 


Finally, when considering the entire strain range between a rearrangement pair, our results indicate that there exists no clear distinction between the rearrangement and its long-range interaction. This suggests the importance of models beyond binary, hysteretic spins in understanding glassy behavior. We note that recent work by Shohat and van Hecke has shown other ways in which spin hysterons models can break down by mapping out the states of interacting, bistable elastic solids~\cite{shohat2024geometric}.

We have shown that rearrangement pairs in jammed systems are inadequately described by binary, spin-like objects; their description needs to be augmented by introducing an energy landscape with double-well potentials. Pairs themselves are common and
tend to exhibit very little hysteresis, features that may help explain why systems are able to fall into periodic orbits with relative ease.  However, this complicates numerical and potentially even experimental studies in which smaller and smaller rearrangement pairs become progressively more difficult to capture. Additional studies are needed to understand the abundance of low-hysteresis rearrangement pairs and their implications for real granular materials. 

\section*{Acknowledgements}

We thank Eddie Bautista, Eric Corwin,  Varda Hagh, Martin van Hecke, Lisa Manning, Damien Vandembroucq, Matthieu Wyart, and Francesco Zamponi for discussions. This work was supported by the US Department of Energy, Office of Science, Basic Energy Sciences, under Grant DE-SC0020972 and the Simons Foundation collaboration,
``Cracking the Glass Problem'', award 348126. C.W.L. was supported in part by NSF Graduate Research Fellowship Grant DGE-1746045 and by the NSF MRSEC program NSF-DMR 2011854.

\section*{Appendix A: application of shear}

The lattice vectors (LV), which define the periodicity of the lattice, are defined as follows:
\begin{equation}
\text{LV} = 
\frac{1}{\sqrt{1 - \gamma^2/4}}
\begin{bmatrix}
1  & \gamma/2\\
\gamma/2 & 1
\end{bmatrix}
\label{eqn:LV}
\end{equation}

Note that this is volume preserving (that is, $\text{det}(\text{LV}) = 1$ for all $\gamma$) and symmetric.

\section*{Appendix B: Detecting rearrangements}

\begin{figure*}
\includegraphics[width=17.2cm]{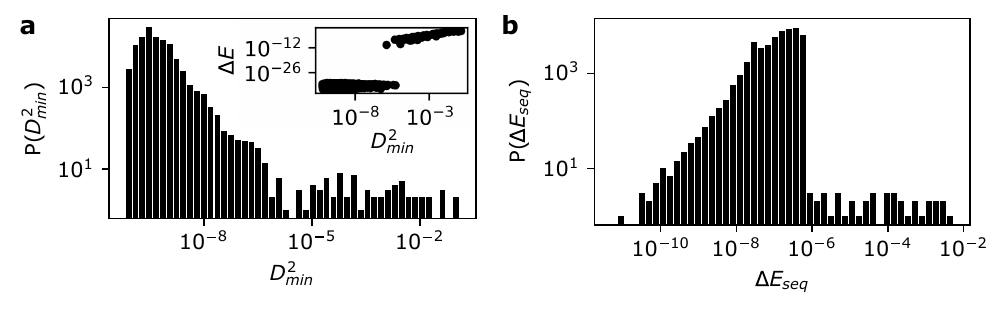}
\caption{For same set of events shown in Fig.~\ref{fig:reversibility}b, (a) distribution of the corresponding $D^2_{min}$ values for sequential frames and (b) distribution of the corresponding change in energy between sequential frames, $\Delta E_{seq}$. Inset to (a) shows $\Delta E$ versus $D^2_{min}$. }
\label{figsi:d2min}
\end{figure*}

We tested three metrics for determining whether a rearrangement has occurred. One is introduced in the main text Fig.~\ref{fig:reversibility}b. Figure~\ref{figsi:d2min} shows the same set of events as in Fig.~\ref{fig:reversibility}b for the two other metrics. The distribution of $D^2_{min}$ values is shown in Fig.~\ref{figsi:d2min}a. We choose as neighbors particles which are within a radius of $1.5 \langle r \rangle$, where $\langle r \rangle$ is the mean particle diameter. Although $D^2_{min}$ is a standard way of determining whether or not a rearrangement has occurred, the distribution reveals no clear distinction between rearrangements and non-rearrangement steps. The inset shows the corresponding $\Delta E$ as a function of $D^2_{min}$; for the same set of events, only $\Delta E$ can distinguish rearrangements unambiguously.

Figure~\ref{figsi:d2min}b shows the equivalent distributions as those in Fig.~\ref{fig:reversibility}b and Fig.~\ref{figsi:d2min} for a third metric: the difference in energies between $\textit{sequential}$ frames (rather than the scheme outlined in the inset to Fig.~1b). 
In both $D^2_{min}$ and $\Delta E_{seq}$, a continuous range of the $x$-axis variable means it is impossible to identify with certainty when a rearrangement has occurred; clearly, $P(|\Delta E|)$ is much better than either of those two measures at discriminating rearrangements from background.

\section*{Appendix C: $N_{core}$ measurements}


The full protocol for determining $N_{core}$ as a function of strain is as follows. Starting from a rearrangement pair with $\gamma^+$ and $\gamma^-$ isolated to a resolution of $10^{-7}$ as described in Methods, the strain distance inside the hysteron was divided linearly into 10 points. For each system and at each strain tested, $N_{core}$ was measured as described in Results and Discussion. Each curve shown is averaged over between 10 and 20 such measurements.

The strain and particle positions are first set from the saved configuration corresponding to the `bottom end' ($\gamma^-$) of the hysteron. A measurement of $N_{core}$ is made as described in the main text, then the strain is incremented to the next of the 10 values measured and the particle positions affinely adjusted before minimizing. Another measurement of $N_{core}$ is made, and so on. We excluded the occasional situations where this procedure caused an artificial shift in $\gamma$ resulting in an early rearrangement.

\begin{figure}
\includegraphics[width=8.6cm]{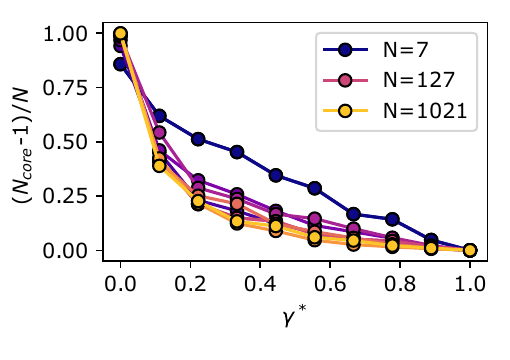}
\caption{Fraction of total particles in the core, ($N_{core}-1)/N$, as a function of strain $\gamma^*$. For clarity, only a few system sizes are shown in the legend; in general, color is as in Fig.~\ref{fig:prob}(a,c) and goes from darkest (small system size) to lightest (large system size).
}
\label{figsi:Ncore}
\end{figure}

In addition to varied hysteresis as reported in the main text, we measured the core size for various $N$ and, aside from data for $N=7$, found no substantial variation as shown in Fig.~\ref{figsi:Ncore}.

\section*{Appendix D: Analysis of $\Delta E$ and $\Delta$ for harmonic and Hertzian springs}

Using a simple spring model for a bistable (double-well) system, we can analytically work out a prediction for scaling behavior between particle displacement $\Delta$, energy drop $\Delta E$, and the hysteresis $\gamma_h$. This is possible for both harmonic and hertzian springs; neither reproduces the results seen the main text. 
Evidently in jammed systems, the second minimum is given by more complex many-particle motion that alters the locally cubic instability, as suggested in the main text.

\begin{figure}
\includegraphics[width=8.6cm]{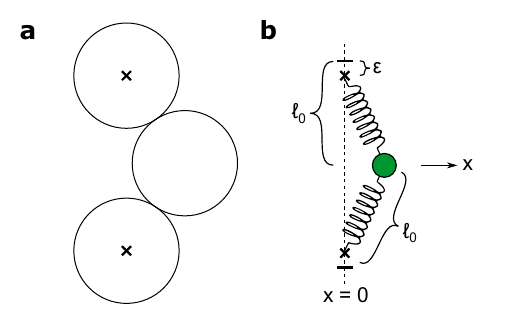}
\caption{(a) Three particle system in which outer particles are fixed and middle particle can move through the gap. (b) Corresponding spring configuration; spring ends are fixed at one end and joined at the green circle. The corresponding energy is a symmetric double well as a function of the x position of the green circle. The distance $\epsilon$ sets positions of the fixed spring ends away from a system with a single minimum at $x=0$. A constant external force along the x direction tilts the energy landscape until one well disappears. Figure adapted from~\cite{lindeman2023competition}.
}
\label{figsi:springs}
\end{figure}

\subsection*{Harmonic potential}

To study an instability analytically, we treat the particles as springs with fixed connectivity and consider the case of a single particle moving through the gap between two others as shown in Fig.~\ref{figsi:springs}a. The resulting spring geometry is shown in Fig.~\ref{figsi:springs}b. For harmonic springs, which store energy as the square of the compression, this corresponds to the simple bistable spring system introduced in~\cite{lindeman2023competition}, with exact energy 
\begin{equation}
    E(x, \gamma) = (\sqrt{(1 - \epsilon)^2 + x^2} - 1)^2,
\end{equation}
where length and force have been non-dimensionalized by the spring's rest length and stiffness.
For small compression $\epsilon$ and sliding motion $x$ this becomes: 
\begin{equation}
    E(x, \gamma) = \frac{x^4}{4} - \epsilon x^2 + \gamma x,
\end{equation}
where $\gamma$ is the external field that pushes the middle particle past the others. This creates a double-well potential which is perfectly symmetric at $\gamma=0$ and becomes unstable at some critical external field value $\pm \gamma_c$.

We can relate the global parameters measured in the main text to features of this spring model: the root mean squared displacement $\Delta$ of a packing is the change in position $\Delta x$ as the system falls from one minimum to the other, the energy drop $\Delta E$ of the packing is the difference in energy between the two wells, and the hysteresis $\gamma_h$ is the external field difference between the left minimum going unstable and the right minimum going unstable so that $\gamma_h = 2 \gamma_c$.

Each of these can be worked out precisely in the spring model. The critical field $\gamma_c$ occurs when the largest gradient in energy (i.e., maximum force) is offset by the external field. This will occur at a value $x_c = \sqrt{2 \epsilon / 3} \sim \epsilon^{1/2}$, determined by setting
\begin{equation}
     -\frac{d^2E}{dx^2} = - 3 x_c^2 + 2\epsilon = 0. 
\end{equation}
The external field $\gamma_c$ that must be applied to precisely balance this force is then 
\begin{equation}
   -\frac{dE}{dx} \Big | _{x=x_c} =  -x_c^3 + 2 \epsilon x_c,
\end{equation}
giving $\gamma_c \sim \epsilon^{3/2}$. Finally, the energy drop associated with this instability is given by the difference in energy at the instability compared with at the minimum. The energy at the instability is just $E(x = x_c, \gamma = \gamma_c) \sim \epsilon^2$. The energy at the minimum can be found to scale the same way, so that the difference $\Delta E \sim \epsilon^2$. The position of this minimum likewise scales the same way as $x_c$ so that $\Delta x \sim x_c \sim \epsilon^{1/2}$. Combining these three results, we see:
$\Delta x \sim \gamma_c^{1/3}$, $\Delta E \sim \gamma_c^{4/3}$, and $\Delta x \sim (\Delta E)^{1/4}$. Recall that $\gamma_c$ in this model is associated with $\gamma_h$ in jammed systems, and $\Delta x$ with $\Delta$. This thus provides a prediction for all three power laws in Fig.~3. These predictions are not consistent with the data.

\subsection*{Hertzian potential}

Above, the argument about the scaling of $\Delta$ and $\Delta E$ with hysteresis made use of a simple spring model with an interaction potential going as overlap squared. Because the simulations reported are for particles with Hertzian interactions, however, we repeat the calculation for springs with $\alpha = 5/2$.

In this case, the energy is 
\begin{equation}
    E(x, \gamma) = (\sqrt{(1 - \epsilon)^2 + x^2} - 1)^{5/2},
\end{equation}
and the expansion for small $\epsilon$ and $x$ is
\begin{equation}
    E(x, \gamma) \approx \frac{15 \sqrt{\epsilon}}{32} x^4 - \frac{5 \epsilon^{3/2}}{4} x^2 + \gamma x.
\end{equation}
Using the same kind of analysis as above, we find the following predictions: $\Delta x \sim \gamma_c^{1/4}$, $\Delta E \sim \gamma_c^{5/4}$, and $\Delta x  \sim (\Delta E)^{1/5}$. Compared with the harmonic results, these are further from the fits shown.

\subsection*{What is held fixed during relaxation}

While the analysis in~\cite{ji2022mean} calculates scaling laws as a function of the gap frequency $\omega_c$ of the glass, the analytic calculations above considered scaling laws as a function of distance $\epsilon$ from the uncompressed state. In a many-particle system, $\epsilon$ itself may effectively change over the course of the rearrangement event due to reorganization of nearby particles, raising the question of which parameters are roughly fixed for a single rearrangement and which must be allowed to vary. 

It is the parameter $\epsilon$ that makes it possible to obtain specific predictions from the calculations above. To see why, consider an argument using a generic quartic like 
$$E = \alpha x^4 - \beta x^2.$$
In this case, one can still calculate the height of the energy barrier and the distance between minima: $\Delta E \sim \beta^2/\alpha$ and $\Delta x \sim \sqrt{\beta/\alpha}$. One could then write either (a) $\Delta E \sim \alpha \Delta x^4$, suggesting $\Delta x \sim (\Delta E)^{1/4}$ or (b) $\Delta E \sim \beta \Delta x^2$, suggesting that $\Delta x \sim (\Delta E)^{1/2}$. Without further argument as to how the coefficients $\alpha$ and $\beta$ are related, this line of thinking does not provide a clear prediction.  The analytic calculations for the two spring models above explicitly include the relationships between $\alpha$ and $\beta$.

\section*{Appendix E: Measured trajectory compared to fold instability analysis}

There is a suggestive similarity between the roughly cubic relationship between $\Delta E$ and $\Delta$ observed in Fig.~\ref{fig:hysteresis} and the ``fold instability" cubic generically obtained when a minimum (lowest order $x^2$) vanishes, leaving $$E(x) \sim \Delta^3.$$
A rearrangement, however, experiences not only this cubic term but also the stabilizing presence of a second well.  This necessitates additional terms. If we ignore those additional terms and concentrate only on the cubic one, our analysis below rules out the possibility that the fold instability \textit{alone} is directly responsible for the relationship measured in Fig.~\ref{fig:hysteresis}b.  That is, other terms rapidly enter to modify this functional form.

\begin{figure*}
\includegraphics[width=17.2cm]{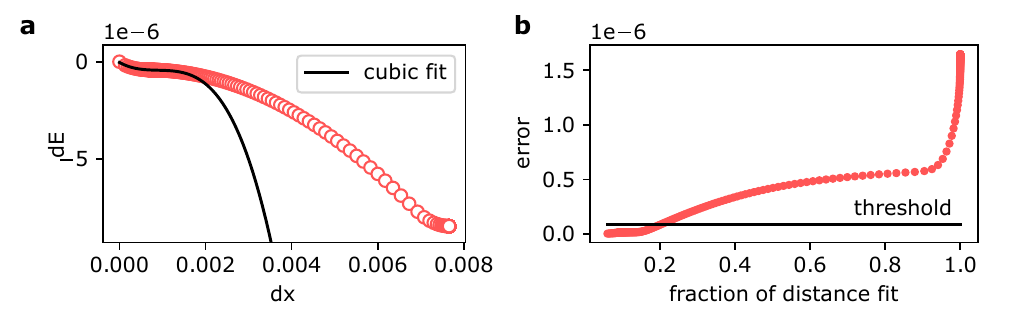}
\caption{(a) The position and energy during minimization over the course of an instability in a 31-particle packing with $\phi = 0.95$ and harmonic contacts. (b) The error of the cubic as a function of the fraction of the distance between minima which was fit. A threshold of $5\%$ above the initial error is shown in black; this value was used to determine how many points to include in the fit shown in (a).}
\label{figsi:dynamics}
\end{figure*}

To assess whether this cubic relation could explain the scaling behavior seen in Fig.~3, we tracked the energy and particle positions during the minimization process as the system transitions out of the marginally-unstable ``minimum''. The process was halted at points evenly spaced in terms of number of minimization steps and the positions and total energy recorded. The energy was then plotted as a function of distance from the initial configuration. A typical resulting ``landscape'', or potential energy change dE versus displacement dx during the minimization, is shown in Fig.~\ref{figsi:dynamics}a.

We first fit a cubic to the 10 data points closest to the initial configuration (leftmost point in the figure), then to the first 11 data points, then 12, and so on. For each fit we computed the total error as the norm of the distance between the data and the fit for each point. As seen in Fig.~\ref{figsi:dynamics}b, this total error is quite low for some fraction of the points, then begins to rise dramatically. By setting a threshold, in this case $5\%$ increase in error, we can estimate how much of the x-range in Fig.~\ref{figsi:dynamics}a this cubic fit represents. Typically, this value was only a small fraction of the total distance from initial configuration to minimized configuration (less than 20$\%$ in the example shown; sometime only a few percent). This emphasizes that the inclusion of only the cubic term determined by the shape of the initial, (marginally) unstable well does not give a good account of the energy landscape throughout the trajectory of relaxation. 

\bibliography{main}

\begin{thebibliography}{40}%
\makeatletter
\providecommand \@ifxundefined [1]{%
 \@ifx{#1\undefined}
}%
\providecommand \@ifnum [1]{%
 \ifnum #1\expandafter \@firstoftwo
 \else \expandafter \@secondoftwo
 \fi
}%
\providecommand \@ifx [1]{%
 \ifx #1\expandafter \@firstoftwo
 \else \expandafter \@secondoftwo
 \fi
}%
\providecommand \natexlab [1]{#1}%
\providecommand \enquote  [1]{``#1''}%
\providecommand \bibnamefont  [1]{#1}%
\providecommand \bibfnamefont [1]{#1}%
\providecommand \citenamefont [1]{#1}%
\providecommand \href@noop [0]{\@secondoftwo}%
\providecommand \href [0]{\begingroup \@sanitize@url \@href}%
\providecommand \@href[1]{\@@startlink{#1}\@@href}%
\providecommand \@@href[1]{\endgroup#1\@@endlink}%
\providecommand \@sanitize@url [0]{\catcode `\\12\catcode `\$12\catcode
  `\&12\catcode `\#12\catcode `\^12\catcode `\_12\catcode `\%12\relax}%
\providecommand \@@startlink[1]{}%
\providecommand \@@endlink[0]{}%
\providecommand \url  [0]{\begingroup\@sanitize@url \@url }%
\providecommand \@url [1]{\endgroup\@href {#1}{\urlprefix }}%
\providecommand \urlprefix  [0]{URL }%
\providecommand \Eprint [0]{\href }%
\providecommand \doibase [0]{http://dx.doi.org/}%
\providecommand \selectlanguage [0]{\@gobble}%
\providecommand \bibinfo  [0]{\@secondoftwo}%
\providecommand \bibfield  [0]{\@secondoftwo}%
\providecommand \translation [1]{[#1]}%
\providecommand \BibitemOpen [0]{}%
\providecommand \bibitemStop [0]{}%
\providecommand \bibitemNoStop [0]{.\EOS\space}%
\providecommand \EOS [0]{\spacefactor3000\relax}%
\providecommand \BibitemShut  [1]{\csname bibitem#1\endcsname}%
\let\auto@bib@innerbib\@empty
\bibitem [{\citenamefont {Falk}\ and\ \citenamefont
  {Langer}(1998)}]{falk1998dynamics}%
  \BibitemOpen
  \bibfield  {author} {\bibinfo {author} {\bibfnamefont {M.~L.}\ \bibnamefont
  {Falk}}\ and\ \bibinfo {author} {\bibfnamefont {J.~S.}\ \bibnamefont
  {Langer}},\ }\href@noop {} {\bibfield  {journal} {\bibinfo  {journal} {Phys.
  Rev. E}\ }\textbf {\bibinfo {volume} {57}},\ \bibinfo {pages} {7192}
  (\bibinfo {year} {1998})}\BibitemShut {NoStop}%
\bibitem [{\citenamefont {Keim}\ and\ \citenamefont
  {Arratia}(2013)}]{keim2013yielding}%
  \BibitemOpen
  \bibfield  {author} {\bibinfo {author} {\bibfnamefont {N.~C.}\ \bibnamefont
  {Keim}}\ and\ \bibinfo {author} {\bibfnamefont {P.~E.}\ \bibnamefont
  {Arratia}},\ }\href@noop {} {\bibfield  {journal} {\bibinfo  {journal} {Soft
  Matter}\ }\textbf {\bibinfo {volume} {9}},\ \bibinfo {pages} {6222} (\bibinfo
  {year} {2013})}\BibitemShut {NoStop}%
\bibitem [{\citenamefont {Regev}\ \emph {et~al.}(2013)\citenamefont {Regev},
  \citenamefont {Lookman},\ and\ \citenamefont {Reichhardt}}]{regev2013onset}%
  \BibitemOpen
  \bibfield  {author} {\bibinfo {author} {\bibfnamefont {I.}~\bibnamefont
  {Regev}}, \bibinfo {author} {\bibfnamefont {T.}~\bibnamefont {Lookman}}, \
  and\ \bibinfo {author} {\bibfnamefont {C.}~\bibnamefont {Reichhardt}},\
  }\href@noop {} {\bibfield  {journal} {\bibinfo  {journal} {Phys. Rev. E}\
  }\textbf {\bibinfo {volume} {88}},\ \bibinfo {pages} {062401} (\bibinfo
  {year} {2013})}\BibitemShut {NoStop}%
\bibitem [{\citenamefont {Fiocco}\ \emph {et~al.}(2013)\citenamefont {Fiocco},
  \citenamefont {Foffi},\ and\ \citenamefont {Sastry}}]{fiocco2013oscillatory}%
  \BibitemOpen
  \bibfield  {author} {\bibinfo {author} {\bibfnamefont {D.}~\bibnamefont
  {Fiocco}}, \bibinfo {author} {\bibfnamefont {G.}~\bibnamefont {Foffi}}, \
  and\ \bibinfo {author} {\bibfnamefont {S.}~\bibnamefont {Sastry}},\
  }\href@noop {} {\bibfield  {journal} {\bibinfo  {journal} {Physical Review
  E}\ }\textbf {\bibinfo {volume} {88}},\ \bibinfo {pages} {020301(R)}
  (\bibinfo {year} {2013})}\BibitemShut {NoStop}%
\bibitem [{\citenamefont {Perchikov}\ and\ \citenamefont
  {Bouchbinder}(2014)}]{perchikov2014variable}%
  \BibitemOpen
  \bibfield  {author} {\bibinfo {author} {\bibfnamefont {N.}~\bibnamefont
  {Perchikov}}\ and\ \bibinfo {author} {\bibfnamefont {E.}~\bibnamefont
  {Bouchbinder}},\ }\href@noop {} {\bibfield  {journal} {\bibinfo  {journal}
  {Physical Review E}\ }\textbf {\bibinfo {volume} {89}},\ \bibinfo {pages}
  {062307} (\bibinfo {year} {2014})}\BibitemShut {NoStop}%
\bibitem [{\citenamefont {Keim}\ and\ \citenamefont
  {Arratia}(2014)}]{keim2014mechanical}%
  \BibitemOpen
  \bibfield  {author} {\bibinfo {author} {\bibfnamefont {N.~C.}\ \bibnamefont
  {Keim}}\ and\ \bibinfo {author} {\bibfnamefont {P.~E.}\ \bibnamefont
  {Arratia}},\ }\href@noop {} {\bibfield  {journal} {\bibinfo  {journal}
  {Physical Review Letters}\ }\textbf {\bibinfo {volume} {112}},\ \bibinfo
  {pages} {028302} (\bibinfo {year} {2014})}\BibitemShut {NoStop}%
\bibitem [{\citenamefont {Fiocco}\ \emph {et~al.}(2014)\citenamefont {Fiocco},
  \citenamefont {Foffi},\ and\ \citenamefont {Sastry}}]{fiocco2014encoding}%
  \BibitemOpen
  \bibfield  {author} {\bibinfo {author} {\bibfnamefont {D.}~\bibnamefont
  {Fiocco}}, \bibinfo {author} {\bibfnamefont {G.}~\bibnamefont {Foffi}}, \
  and\ \bibinfo {author} {\bibfnamefont {S.}~\bibnamefont {Sastry}},\
  }\href@noop {} {\bibfield  {journal} {\bibinfo  {journal} {Phys. Rev.
  Letts.}\ }\textbf {\bibinfo {volume} {112}},\ \bibinfo {pages} {025702}
  (\bibinfo {year} {2014})}\BibitemShut {NoStop}%
\bibitem [{\citenamefont {Royer}\ and\ \citenamefont
  {Chaikin}(2015)}]{royer2015precisely}%
  \BibitemOpen
  \bibfield  {author} {\bibinfo {author} {\bibfnamefont {J.~R.}\ \bibnamefont
  {Royer}}\ and\ \bibinfo {author} {\bibfnamefont {P.~M.}\ \bibnamefont
  {Chaikin}},\ }\href@noop {} {\bibfield  {journal} {\bibinfo  {journal}
  {Proceedings of the National Academy of Sciences}\ }\textbf {\bibinfo
  {volume} {112}},\ \bibinfo {pages} {49} (\bibinfo {year} {2015})}\BibitemShut
  {NoStop}%
\bibitem [{\citenamefont {Lavrentovich}\ \emph {et~al.}(2017)\citenamefont
  {Lavrentovich}, \citenamefont {Liu},\ and\ \citenamefont
  {Nagel}}]{lavrentovich2017period}%
  \BibitemOpen
  \bibfield  {author} {\bibinfo {author} {\bibfnamefont {M.~O.}\ \bibnamefont
  {Lavrentovich}}, \bibinfo {author} {\bibfnamefont {A.~J.}\ \bibnamefont
  {Liu}}, \ and\ \bibinfo {author} {\bibfnamefont {S.~R.}\ \bibnamefont
  {Nagel}},\ }\href@noop {} {\bibfield  {journal} {\bibinfo  {journal} {Phys.
  Rev. E}\ }\textbf {\bibinfo {volume} {96}},\ \bibinfo {pages} {020101(R)}
  (\bibinfo {year} {2017})}\BibitemShut {NoStop}%
\bibitem [{\citenamefont {Maloney}\ and\ \citenamefont
  {Lemaitre}(2004{\natexlab{a}})}]{maloney2004subextensive}%
  \BibitemOpen
  \bibfield  {author} {\bibinfo {author} {\bibfnamefont {C.}~\bibnamefont
  {Maloney}}\ and\ \bibinfo {author} {\bibfnamefont {A.}~\bibnamefont
  {Lemaitre}},\ }\href@noop {} {\bibfield  {journal} {\bibinfo  {journal}
  {Physical Review Letters}\ }\textbf {\bibinfo {volume} {93}},\ \bibinfo
  {pages} {016001} (\bibinfo {year} {2004}{\natexlab{a}})}\BibitemShut
  {NoStop}%
\bibitem [{\citenamefont {Maloney}\ and\ \citenamefont
  {Lemaitre}(2004{\natexlab{b}})}]{maloney2004universal}%
  \BibitemOpen
  \bibfield  {author} {\bibinfo {author} {\bibfnamefont {C.}~\bibnamefont
  {Maloney}}\ and\ \bibinfo {author} {\bibfnamefont {A.}~\bibnamefont
  {Lemaitre}},\ }\href@noop {} {\bibfield  {journal} {\bibinfo  {journal}
  {Physical Review Letters}\ }\textbf {\bibinfo {volume} {93}},\ \bibinfo
  {pages} {195501} (\bibinfo {year} {2004}{\natexlab{b}})}\BibitemShut
  {NoStop}%
\bibitem [{\citenamefont {Maloney}\ and\ \citenamefont
  {Lemaitre}(2006)}]{maloney2006amorphous}%
  \BibitemOpen
  \bibfield  {author} {\bibinfo {author} {\bibfnamefont {C.~E.}\ \bibnamefont
  {Maloney}}\ and\ \bibinfo {author} {\bibfnamefont {A.}~\bibnamefont
  {Lemaitre}},\ }\href@noop {} {\bibfield  {journal} {\bibinfo  {journal}
  {Physical Review E—Statistical, Nonlinear, and Soft Matter Physics}\
  }\textbf {\bibinfo {volume} {74}},\ \bibinfo {pages} {016118} (\bibinfo
  {year} {2006})}\BibitemShut {NoStop}%
\bibitem [{\citenamefont {Shimada}\ \emph {et~al.}(2018)\citenamefont
  {Shimada}, \citenamefont {Mizuno}, \citenamefont {Wyart},\ and\ \citenamefont
  {Ikeda}}]{shimada2018spatial}%
  \BibitemOpen
  \bibfield  {author} {\bibinfo {author} {\bibfnamefont {M.}~\bibnamefont
  {Shimada}}, \bibinfo {author} {\bibfnamefont {H.}~\bibnamefont {Mizuno}},
  \bibinfo {author} {\bibfnamefont {M.}~\bibnamefont {Wyart}}, \ and\ \bibinfo
  {author} {\bibfnamefont {A.}~\bibnamefont {Ikeda}},\ }\href@noop {}
  {\bibfield  {journal} {\bibinfo  {journal} {Physical Review E}\ }\textbf
  {\bibinfo {volume} {98}},\ \bibinfo {pages} {060901} (\bibinfo {year}
  {2018})}\BibitemShut {NoStop}%
\bibitem [{\citenamefont {Xu}\ \emph {et~al.}(2017)\citenamefont {Xu},
  \citenamefont {Liu},\ and\ \citenamefont {Nagel}}]{xu2017instabilities}%
  \BibitemOpen
  \bibfield  {author} {\bibinfo {author} {\bibfnamefont {N.}~\bibnamefont
  {Xu}}, \bibinfo {author} {\bibfnamefont {A.~J.}\ \bibnamefont {Liu}}, \ and\
  \bibinfo {author} {\bibfnamefont {S.~R.}\ \bibnamefont {Nagel}},\ }\href@noop
  {} {\bibfield  {journal} {\bibinfo  {journal} {Physical Review Letters}\
  }\textbf {\bibinfo {volume} {119}},\ \bibinfo {pages} {215502} (\bibinfo
  {year} {2017})}\BibitemShut {NoStop}%
\bibitem [{\citenamefont {Ruan}\ \emph {et~al.}(2022)\citenamefont {Ruan},
  \citenamefont {Patinet},\ and\ \citenamefont {Falk}}]{ruan2022predicting}%
  \BibitemOpen
  \bibfield  {author} {\bibinfo {author} {\bibfnamefont {D.}~\bibnamefont
  {Ruan}}, \bibinfo {author} {\bibfnamefont {S.}~\bibnamefont {Patinet}}, \
  and\ \bibinfo {author} {\bibfnamefont {M.~L.}\ \bibnamefont {Falk}},\
  }\href@noop {} {\bibfield  {journal} {\bibinfo  {journal} {Journal of the
  Mechanics and Physics of Solids}\ }\textbf {\bibinfo {volume} {158}},\
  \bibinfo {pages} {104671} (\bibinfo {year} {2022})}\BibitemShut {NoStop}%
\bibitem [{\citenamefont {Xu}\ \emph {et~al.}(2023)\citenamefont {Xu},
  \citenamefont {Zhang}, \citenamefont {Liu}, \citenamefont {Nagel},\ and\
  \citenamefont {Xu}}]{xu2023instabilities}%
  \BibitemOpen
  \bibfield  {author} {\bibinfo {author} {\bibfnamefont {D.}~\bibnamefont
  {Xu}}, \bibinfo {author} {\bibfnamefont {S.}~\bibnamefont {Zhang}}, \bibinfo
  {author} {\bibfnamefont {A.~J.}\ \bibnamefont {Liu}}, \bibinfo {author}
  {\bibfnamefont {S.~R.}\ \bibnamefont {Nagel}}, \ and\ \bibinfo {author}
  {\bibfnamefont {N.}~\bibnamefont {Xu}},\ }\href@noop {} {\bibfield  {journal}
  {\bibinfo  {journal} {Proceedings of the National Academy of Sciences}\
  }\textbf {\bibinfo {volume} {120}},\ \bibinfo {pages} {e2304974120} (\bibinfo
  {year} {2023})}\BibitemShut {NoStop}%
\bibitem [{\citenamefont {Preisach}(1935)}]{preisach1935magnetische}%
  \BibitemOpen
  \bibfield  {author} {\bibinfo {author} {\bibfnamefont {F.}~\bibnamefont
  {Preisach}},\ }\href@noop {} {\bibfield  {journal} {\bibinfo  {journal}
  {Zeitschrift f{\"u}r Physik}\ }\textbf {\bibinfo {volume} {94}},\ \bibinfo
  {pages} {277} (\bibinfo {year} {1935})}\BibitemShut {NoStop}%
\bibitem [{\citenamefont {Sethna}\ \emph {et~al.}(1993)\citenamefont {Sethna},
  \citenamefont {Dahmen}, \citenamefont {Kartha}, \citenamefont {Krumhansl},
  \citenamefont {Roberts},\ and\ \citenamefont {Shore}}]{sethna1993hysteresis}%
  \BibitemOpen
  \bibfield  {author} {\bibinfo {author} {\bibfnamefont {J.~P.}\ \bibnamefont
  {Sethna}}, \bibinfo {author} {\bibfnamefont {K.}~\bibnamefont {Dahmen}},
  \bibinfo {author} {\bibfnamefont {S.}~\bibnamefont {Kartha}}, \bibinfo
  {author} {\bibfnamefont {J.~A.}\ \bibnamefont {Krumhansl}}, \bibinfo {author}
  {\bibfnamefont {B.~W.}\ \bibnamefont {Roberts}}, \ and\ \bibinfo {author}
  {\bibfnamefont {J.~D.}\ \bibnamefont {Shore}},\ }\href@noop {} {\bibfield
  {journal} {\bibinfo  {journal} {Phys. Rev. Letts.}\ }\textbf {\bibinfo
  {volume} {70}},\ \bibinfo {pages} {3347} (\bibinfo {year}
  {1993})}\BibitemShut {NoStop}%
\bibitem [{\citenamefont {Mungan}\ \emph {et~al.}(2019)\citenamefont {Mungan},
  \citenamefont {Sastry}, \citenamefont {Dahmen},\ and\ \citenamefont
  {Regev}}]{mungan2019networks}%
  \BibitemOpen
  \bibfield  {author} {\bibinfo {author} {\bibfnamefont {M.}~\bibnamefont
  {Mungan}}, \bibinfo {author} {\bibfnamefont {S.}~\bibnamefont {Sastry}},
  \bibinfo {author} {\bibfnamefont {K.}~\bibnamefont {Dahmen}}, \ and\ \bibinfo
  {author} {\bibfnamefont {I.}~\bibnamefont {Regev}},\ }\href@noop {}
  {\bibfield  {journal} {\bibinfo  {journal} {Phys. Rev. Letts.}\ }\textbf
  {\bibinfo {volume} {123}},\ \bibinfo {pages} {178002} (\bibinfo {year}
  {2019})}\BibitemShut {NoStop}%
\bibitem [{\citenamefont {Keim}\ \emph {et~al.}(2020)\citenamefont {Keim},
  \citenamefont {Hass}, \citenamefont {Kroger},\ and\ \citenamefont
  {Wieker}}]{keim2020global}%
  \BibitemOpen
  \bibfield  {author} {\bibinfo {author} {\bibfnamefont {N.~C.}\ \bibnamefont
  {Keim}}, \bibinfo {author} {\bibfnamefont {J.}~\bibnamefont {Hass}}, \bibinfo
  {author} {\bibfnamefont {B.}~\bibnamefont {Kroger}}, \ and\ \bibinfo {author}
  {\bibfnamefont {D.}~\bibnamefont {Wieker}},\ }\href@noop {} {\bibfield
  {journal} {\bibinfo  {journal} {Phys. Rev. Research}\ }\textbf {\bibinfo
  {volume} {2}},\ \bibinfo {pages} {012004(R)} (\bibinfo {year}
  {2020})}\BibitemShut {NoStop}%
\bibitem [{\citenamefont {Terzi}\ and\ \citenamefont
  {Mungan}(2020)}]{terzi2020state}%
  \BibitemOpen
  \bibfield  {author} {\bibinfo {author} {\bibfnamefont {M.~M.}\ \bibnamefont
  {Terzi}}\ and\ \bibinfo {author} {\bibfnamefont {M.}~\bibnamefont {Mungan}},\
  }\href@noop {} {\bibfield  {journal} {\bibinfo  {journal} {Phys. Rev. E}\
  }\textbf {\bibinfo {volume} {102}},\ \bibinfo {pages} {012122} (\bibinfo
  {year} {2020})}\BibitemShut {NoStop}%
\bibitem [{\citenamefont {van Hecke}(2021)}]{van2021profusion}%
  \BibitemOpen
  \bibfield  {author} {\bibinfo {author} {\bibfnamefont {M.}~\bibnamefont {van
  Hecke}},\ }\href@noop {} {\bibfield  {journal} {\bibinfo  {journal} {Phys.
  Rev. E}\ }\textbf {\bibinfo {volume} {104}},\ \bibinfo {pages} {054608}
  (\bibinfo {year} {2021})}\BibitemShut {NoStop}%
\bibitem [{\citenamefont {Szulc}\ \emph {et~al.}(2022)\citenamefont {Szulc},
  \citenamefont {Mungan},\ and\ \citenamefont {Regev}}]{szulc2022cooperative}%
  \BibitemOpen
  \bibfield  {author} {\bibinfo {author} {\bibfnamefont {A.}~\bibnamefont
  {Szulc}}, \bibinfo {author} {\bibfnamefont {M.}~\bibnamefont {Mungan}}, \
  and\ \bibinfo {author} {\bibfnamefont {I.}~\bibnamefont {Regev}},\
  }\href@noop {} {\bibfield  {journal} {\bibinfo  {journal} {J. Chem. Phys.}\
  }\textbf {\bibinfo {volume} {156}} (\bibinfo {year} {2022})}\BibitemShut
  {NoStop}%
\bibitem [{\citenamefont {Shohat}\ \emph {et~al.}(2022)\citenamefont {Shohat},
  \citenamefont {Hexner},\ and\ \citenamefont {Lahini}}]{shohat2022memory}%
  \BibitemOpen
  \bibfield  {author} {\bibinfo {author} {\bibfnamefont {D.}~\bibnamefont
  {Shohat}}, \bibinfo {author} {\bibfnamefont {D.}~\bibnamefont {Hexner}}, \
  and\ \bibinfo {author} {\bibfnamefont {Y.}~\bibnamefont {Lahini}},\
  }\href@noop {} {\bibfield  {journal} {\bibinfo  {journal} {Proceedings of the
  National Academy of Sciences}\ }\textbf {\bibinfo {volume} {119}},\ \bibinfo
  {pages} {e2200028119} (\bibinfo {year} {2022})}\BibitemShut {NoStop}%
\bibitem [{\citenamefont {Bense}\ and\ \citenamefont {van
  Hecke}(2021)}]{bense2021complex}%
  \BibitemOpen
  \bibfield  {author} {\bibinfo {author} {\bibfnamefont {H.}~\bibnamefont
  {Bense}}\ and\ \bibinfo {author} {\bibfnamefont {M.}~\bibnamefont {van
  Hecke}},\ }\href@noop {} {\bibfield  {journal} {\bibinfo  {journal}
  {Proceedings of the National Academy of Sciences}\ }\textbf {\bibinfo
  {volume} {118}},\ \bibinfo {pages} {e2111436118} (\bibinfo {year}
  {2021})}\BibitemShut {NoStop}%
\bibitem [{\citenamefont {Mungan}(2022)}]{mungan2022putting}%
  \BibitemOpen
  \bibfield  {author} {\bibinfo {author} {\bibfnamefont {M.}~\bibnamefont
  {Mungan}},\ }\href@noop {} {\bibfield  {journal} {\bibinfo  {journal}
  {Proceedings of the National Academy of Sciences}\ }\textbf {\bibinfo
  {volume} {119}},\ \bibinfo {pages} {e2208743119} (\bibinfo {year}
  {2022})}\BibitemShut {NoStop}%
\bibitem [{\citenamefont {Keim}\ and\ \citenamefont
  {Paulsen}(2021)}]{keim2021multiperiodic}%
  \BibitemOpen
  \bibfield  {author} {\bibinfo {author} {\bibfnamefont {N.~C.}\ \bibnamefont
  {Keim}}\ and\ \bibinfo {author} {\bibfnamefont {J.~D.}\ \bibnamefont
  {Paulsen}},\ }\href@noop {} {\bibfield  {journal} {\bibinfo  {journal}
  {Science Advances}\ }\textbf {\bibinfo {volume} {7}},\ \bibinfo {pages}
  {eabg7685} (\bibinfo {year} {2021})}\BibitemShut {NoStop}%
\bibitem [{\citenamefont {Lindeman}\ and\ \citenamefont
  {Nagel}(2021)}]{lindeman2021multiple}%
  \BibitemOpen
  \bibfield  {author} {\bibinfo {author} {\bibfnamefont {C.~W.}\ \bibnamefont
  {Lindeman}}\ and\ \bibinfo {author} {\bibfnamefont {S.~R.}\ \bibnamefont
  {Nagel}},\ }\href@noop {} {\bibfield  {journal} {\bibinfo  {journal} {Science
  Advances}\ }\textbf {\bibinfo {volume} {7}},\ \bibinfo {pages} {eabg7133}
  (\bibinfo {year} {2021})}\BibitemShut {NoStop}%
\bibitem [{\citenamefont {Kumar}\ \emph {et~al.}(2022)\citenamefont {Kumar},
  \citenamefont {Patinet}, \citenamefont {Maloney}, \citenamefont {Regev},
  \citenamefont {Vandembroucq},\ and\ \citenamefont
  {Mungan}}]{kumar2022mapping}%
  \BibitemOpen
  \bibfield  {author} {\bibinfo {author} {\bibfnamefont {D.}~\bibnamefont
  {Kumar}}, \bibinfo {author} {\bibfnamefont {S.}~\bibnamefont {Patinet}},
  \bibinfo {author} {\bibfnamefont {C.~E.}\ \bibnamefont {Maloney}}, \bibinfo
  {author} {\bibfnamefont {I.}~\bibnamefont {Regev}}, \bibinfo {author}
  {\bibfnamefont {D.}~\bibnamefont {Vandembroucq}}, \ and\ \bibinfo {author}
  {\bibfnamefont {M.}~\bibnamefont {Mungan}},\ }\href@noop {} {\bibfield
  {journal} {\bibinfo  {journal} {The Journal of Chemical Physics}\ }\textbf
  {\bibinfo {volume} {157}} (\bibinfo {year} {2022})}\BibitemShut {NoStop}%
\bibitem [{\citenamefont {Morse}\ and\ \citenamefont
  {Corwin}(2014)}]{morse2014geometric}%
  \BibitemOpen
  \bibfield  {author} {\bibinfo {author} {\bibfnamefont {P.~K.}\ \bibnamefont
  {Morse}}\ and\ \bibinfo {author} {\bibfnamefont {E.~I.}\ \bibnamefont
  {Corwin}},\ }\href@noop {} {\bibfield  {journal} {\bibinfo  {journal}
  {Physical Review Letters}\ }\textbf {\bibinfo {volume} {112}},\ \bibinfo
  {pages} {115701} (\bibinfo {year} {2014})}\BibitemShut {NoStop}%
\bibitem [{\citenamefont {Charbonneau}\ \emph {et~al.}(2015)\citenamefont
  {Charbonneau}, \citenamefont {Corwin}, \citenamefont {Parisi},\ and\
  \citenamefont {Zamponi}}]{charbonneau2015jamming}%
  \BibitemOpen
  \bibfield  {author} {\bibinfo {author} {\bibfnamefont {P.}~\bibnamefont
  {Charbonneau}}, \bibinfo {author} {\bibfnamefont {E.~I.}\ \bibnamefont
  {Corwin}}, \bibinfo {author} {\bibfnamefont {G.}~\bibnamefont {Parisi}}, \
  and\ \bibinfo {author} {\bibfnamefont {F.}~\bibnamefont {Zamponi}},\
  }\href@noop {} {\bibfield  {journal} {\bibinfo  {journal} {Physical Review
  Letters}\ }\textbf {\bibinfo {volume} {114}},\ \bibinfo {pages} {125504}
  (\bibinfo {year} {2015})}\BibitemShut {NoStop}%
\bibitem [{\citenamefont {Salerno}\ and\ \citenamefont
  {Robbins}(2013)}]{salerno2013effect}%
  \BibitemOpen
  \bibfield  {author} {\bibinfo {author} {\bibfnamefont {K.~M.}\ \bibnamefont
  {Salerno}}\ and\ \bibinfo {author} {\bibfnamefont {M.~O.}\ \bibnamefont
  {Robbins}},\ }\href@noop {} {\bibfield  {journal} {\bibinfo  {journal}
  {Physical Review E—Statistical, Nonlinear, and Soft Matter Physics}\
  }\textbf {\bibinfo {volume} {88}},\ \bibinfo {pages} {062206} (\bibinfo
  {year} {2013})}\BibitemShut {NoStop}%
\bibitem [{\citenamefont {Sastry}\ \emph {et~al.}(1998)\citenamefont {Sastry},
  \citenamefont {Debenedetti},\ and\ \citenamefont
  {Stillinger}}]{sastry1998signatures}%
  \BibitemOpen
  \bibfield  {author} {\bibinfo {author} {\bibfnamefont {S.}~\bibnamefont
  {Sastry}}, \bibinfo {author} {\bibfnamefont {P.~G.}\ \bibnamefont
  {Debenedetti}}, \ and\ \bibinfo {author} {\bibfnamefont {F.~H.}\ \bibnamefont
  {Stillinger}},\ }\href@noop {} {\bibfield  {journal} {\bibinfo  {journal}
  {Nature}\ }\textbf {\bibinfo {volume} {393}},\ \bibinfo {pages} {554}
  (\bibinfo {year} {1998})}\BibitemShut {NoStop}%
\bibitem [{\citenamefont {Silbert}\ \emph {et~al.}(2009)\citenamefont
  {Silbert}, \citenamefont {Liu},\ and\ \citenamefont
  {Nagel}}]{silbert2009normal}%
  \BibitemOpen
  \bibfield  {author} {\bibinfo {author} {\bibfnamefont {L.~E.}\ \bibnamefont
  {Silbert}}, \bibinfo {author} {\bibfnamefont {A.~J.}\ \bibnamefont {Liu}}, \
  and\ \bibinfo {author} {\bibfnamefont {S.~R.}\ \bibnamefont {Nagel}},\
  }\href@noop {} {\bibfield  {journal} {\bibinfo  {journal} {Physical Review
  E}\ }\textbf {\bibinfo {volume} {79}},\ \bibinfo {pages} {021308} (\bibinfo
  {year} {2009})}\BibitemShut {NoStop}%
\bibitem [{SI-()}]{SI-ref}%
  \BibitemOpen
  \href@noop {} {\bibinfo  {journal} {See Supplemental Material at [URL will be
  inserted by publisher] for definition of boundary conditions; detection of
  rearrangement events; analytical prediction of scaling between particle
  displacements, energy drop, and hysteresis during an instability; scaling
  data for harmonic spheres, higher density packings, and larger system size;
  and details of reported measurements}\ }\BibitemShut {NoStop}%
\bibitem [{\citenamefont {Lindeman}\ \emph {et~al.}(2023)\citenamefont
  {Lindeman}, \citenamefont {Hagh}, \citenamefont {Ip},\ and\ \citenamefont
  {Nagel}}]{lindeman2023competition}%
  \BibitemOpen
\bibfield  {journal} {  }\bibfield  {author} {\bibinfo {author} {\bibfnamefont
  {C.~W.}\ \bibnamefont {Lindeman}}, \bibinfo {author} {\bibfnamefont {V.~F.}\
  \bibnamefont {Hagh}}, \bibinfo {author} {\bibfnamefont {C.~I.}\ \bibnamefont
  {Ip}}, \ and\ \bibinfo {author} {\bibfnamefont {S.~R.}\ \bibnamefont
  {Nagel}},\ }\href@noop {} {\bibfield  {journal} {\bibinfo  {journal} {Phys.
  Rev. Letts.}\ }\textbf {\bibinfo {volume} {130}},\ \bibinfo {pages} {197201}
  (\bibinfo {year} {2023})}\BibitemShut {NoStop}%
\bibitem [{\citenamefont {Gao}\ \emph {et~al.}(2009)\citenamefont {Gao},
  \citenamefont {Blawzdziewicz},\ and\ \citenamefont {O'Hern}}]{gao2009}%
  \BibitemOpen
  \bibfield  {author} {\bibinfo {author} {\bibfnamefont {G.-J.}\ \bibnamefont
  {Gao}}, \bibinfo {author} {\bibfnamefont {J.}~\bibnamefont {Blawzdziewicz}},
  \ and\ \bibinfo {author} {\bibfnamefont {C.~S.}\ \bibnamefont {O'Hern}},\
  }\href@noop {} {\bibfield  {journal} {\bibinfo  {journal} {Phys. Rev. E.}\
  }\textbf {\bibinfo {volume} {80}},\ \bibinfo {pages} {1061303} (\bibinfo
  {year} {2009})}\BibitemShut {NoStop}%
\bibitem [{\citenamefont {Hagh}\ and\ \citenamefont {Nagel}(2024)}]{hagh2024}%
  \BibitemOpen
  \bibfield  {author} {\bibinfo {author} {\bibfnamefont {V.~F.}\ \bibnamefont
  {Hagh}}\ and\ \bibinfo {author} {\bibfnamefont {S.~R.}\ \bibnamefont
  {Nagel}},\ }\href@noop {} {\bibfield  {journal} {\bibinfo  {journal}
  {preprint arXiv:2403.03926}\ } (\bibinfo {year} {2024})}\BibitemShut
  {NoStop}%
\bibitem [{\citenamefont {Ji}\ \emph {et~al.}(2022)\citenamefont {Ji},
  \citenamefont {de~Geus}, \citenamefont {Agoritsas},\ and\ \citenamefont
  {Wyart}}]{ji2022mean}%
  \BibitemOpen
  \bibfield  {author} {\bibinfo {author} {\bibfnamefont {W.}~\bibnamefont
  {Ji}}, \bibinfo {author} {\bibfnamefont {T.~W.}\ \bibnamefont {de~Geus}},
  \bibinfo {author} {\bibfnamefont {E.}~\bibnamefont {Agoritsas}}, \ and\
  \bibinfo {author} {\bibfnamefont {M.}~\bibnamefont {Wyart}},\ }\href@noop {}
  {\bibfield  {journal} {\bibinfo  {journal} {Physical Review E}\ }\textbf
  {\bibinfo {volume} {105}},\ \bibinfo {pages} {044601} (\bibinfo {year}
  {2022})}\BibitemShut {NoStop}%
\bibitem [{\citenamefont {Shohat}\ and\ \citenamefont {van
  Hecke}(2024)}]{shohat2024geometric}%
  \BibitemOpen
  \bibfield  {author} {\bibinfo {author} {\bibfnamefont {D.}~\bibnamefont
  {Shohat}}\ and\ \bibinfo {author} {\bibfnamefont {M.}~\bibnamefont {van
  Hecke}},\ }\href@noop {} {\bibfield  {journal} {\bibinfo  {journal} {arXiv
  preprint arXiv:2409.07804}\ } (\bibinfo {year} {2024})}\BibitemShut {NoStop}%
\end{thebibliography}%

\end{document}


\renewcommand{\thefigure}{S\arabic{figure}}
\setcounter{figure}{0} 
\renewcommand{\theequation}{S\arabic{equation}}
\setcounter{equation}{0}
\renewcommand{\thetable}{S\arabic{table}}
\setcounter{table}{0}

\title{Supplementary Material for \\ ``Minimal cyclic behavior in sheared amorphous solids''}

\date{\today}

\maketitle

\renewcommand{\thefigure}{S\arabic{figure}}
\setcounter{figure}{0} 
\renewcommand{\theequation}{S\arabic{equation}}
\setcounter{equation}{0}

\subsection*{Additional scaling details}

\textit{Particle displacements --- }
The root-mean-square particle displacement $\Delta$, defined in the main text, was calculated after mapping the particles back to a square box configuration using the inverse of the lattice vectors. Particles which were rattlers (ie, have no stable contacts) before and/or after the rearrangement were excluded from the calculation and the drift, or average motion of all the particles, was subtracted off.

\begin{figure*}
\includegraphics[width=8.6cm]{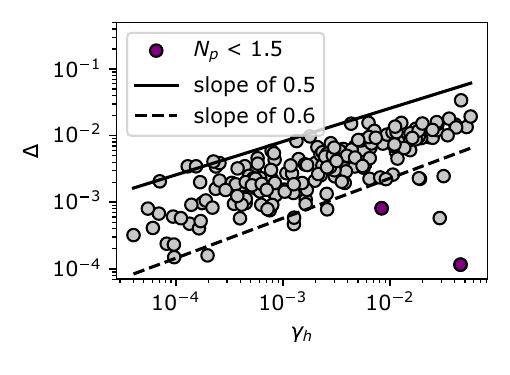}
\caption{RMS displacements $\Delta$ of particles as a function of the hysteresis between two paired instabilities $\gamma_h$ for the same instabilities shown in Fig.~3 in the main text ($N=31$ and $\phi = 0.95$).}
\label{figsi:powerlaw}
\end{figure*}

\textit{$\Delta$ vs $\gamma_h$ --- }
Here we show additional results on the correlation between $\Delta$, $\Delta E$, and $\gamma_h$. The relation between $\Delta$ and $\gamma_h$ is shown in Fig.~\ref{figsi:powerlaw} for the parameters used in the main text Fig.~3. Slope estimates 0.5 to 0.6 are as expected based on the product of the slopes shown in Fig.~3a and Fig.~3b.

\begin{figure*}
\includegraphics[width=17.2cm]{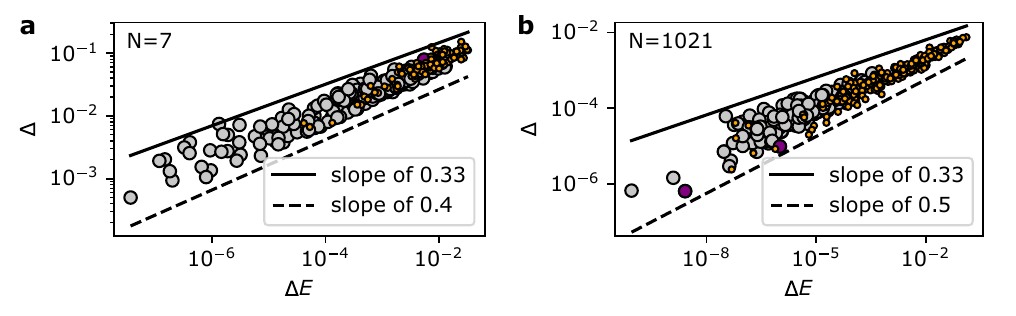}
\caption{Scaling between displacements $\Delta$ and energy drop $\Delta E$ for $\phi=0.95$ and (a) $N=7$ and (b) N=1021. Slopes shown are guides to the eye.}
\label{figsi:large}
\end{figure*}

\textit{System size --- }
Figure~\ref{figsi:large} shows displacements $\Delta$ and energy drop $\Delta E$ for the smallest and largest system sizes studied, $N=7$ and $N=1021$. Though the bounds shown vary slightly from those in the main text, no systematic trend was apparent across system sizes.

\begin{figure}
\includegraphics[width=17.2cm]{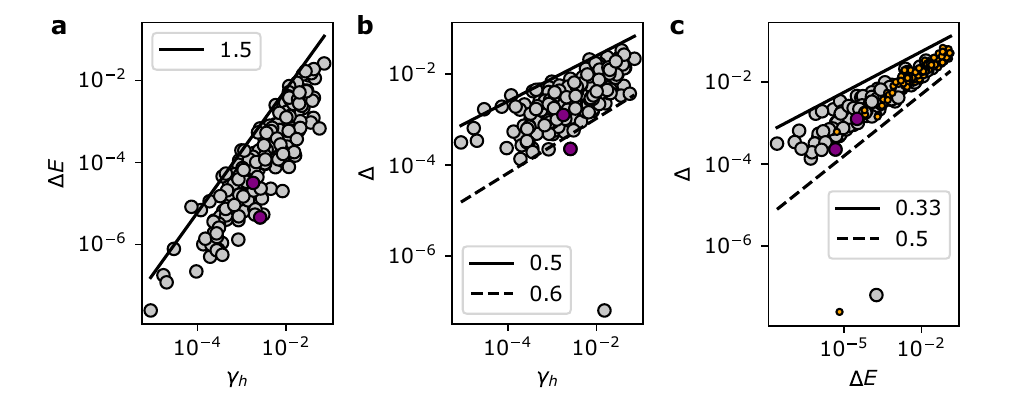}
\caption{Scaling between different features associated with rearrangements in harmonic packings ($N=31$ and $\phi=0.99$).  Lines are guides to the eye intended to bound the data. (a) Energy drop and (b) root-mean-square displacement $\Delta$ for hysteron rearrangements measured at $\gamma^+$ versus hysteresis $\gamma_h$. (c) Root-mean-square displacements $\Delta$ versus energy drop $\Delta E$ for the same rearrangements. The scaling behaviors are not statistically different from those for Hertzian packings.}
\label{figsi:harmonic}
\end{figure}

\textit{Harmonic interactions --- }
Scalings for $N=31$ with harmonic interactions between particles ($\phi = 0.95$). The resulting $\Delta$, $\Delta E$, and $\gamma_h$ plots are shown in Fig.~\ref{figsi:harmonic}. 

\subsection*{Participation ratio}

If $r_i$ is the distance each particle moves, the participation ratio of the motion is defined as 
\begin{equation}
P = \frac{(\Sigma_i r_i^2 )^2}{\Sigma_i r_i^4},
\end{equation} a quantity that lies between $1/N$ and $1$. The participation number $N_p$ noted in the main text is then $N_p = P*N$.

\begin{figure}
\includegraphics[width=17.2cm]{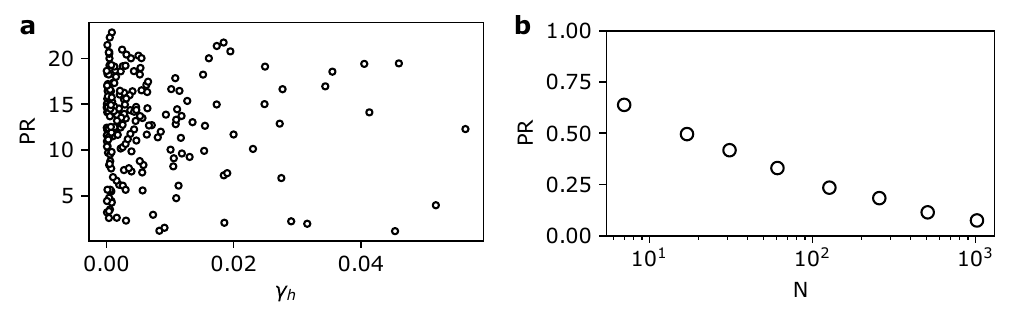}
\caption{(a) The participation ratio of paired rearrangements as a function of hysteresis ($N=31$ and $\phi=0.95$). (b) The average participation ratio of the ``top'' of paired rearrangements as a function of system size ($\phi=0.95$).  }
\label{figsi:PR}
\end{figure}

Figure~\ref{figsi:PR}a shows the participation ratio 
as a function of 
hysteresis. The participation ratio 
shows no consistent trend over more than two orders of magnitude in hysteresis. This supports the claim in the main text that the participation ratio of rearrangements in a pair does not depend in a systematic way on the hysteresis of the pair. 

The participation ratio does depend on the system size: it seems to fall roughly logarithmically with $N$. This trend cannot continue forever; in the limit of large $N$, the participation ratio must plateau to zero (the beginning of this plateau appears to be visible for the largest system sizes studied).

\begin{figure}
\includegraphics[width=7.5cm]{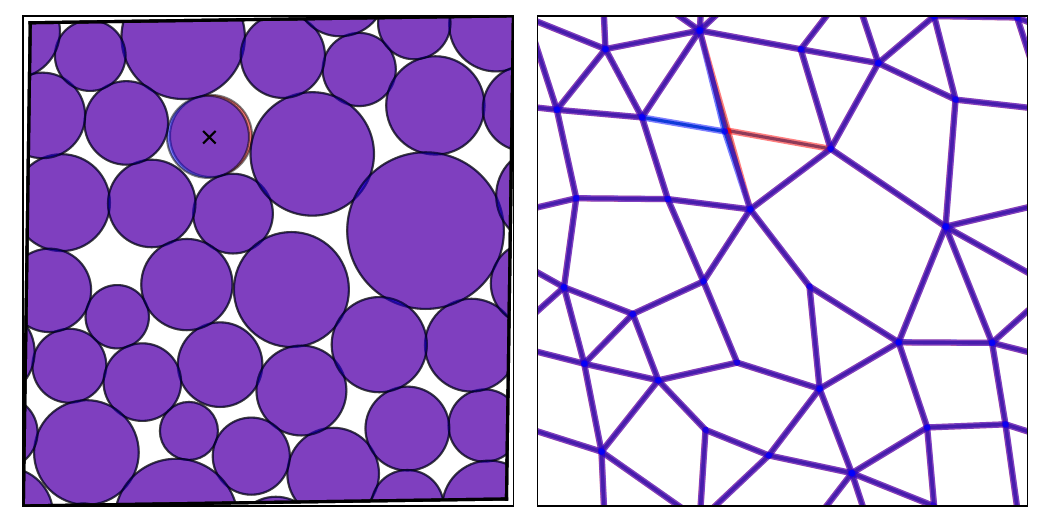}
\caption{Example of a rearrangement with low participation number, $N_p < 1.5$, with motion dominated by a buckler. Left: particle packing shown directly; red and blue show the configurations of the particles before and after the rearrangement, respectively, with most of the packing unchanged by the rearrangement and hence purple. Slivers of red and blue are just visible on either side of the buckler, which is marked with an x. Right: the associated contact network, colored as in (a), showing a single change in contact associated with the buckler buckling from left to right.}
\label{figsi:bucklers}
\end{figure}

\textit{Bucklers} --- Rearrangements with participation number close to 1 (that is, rearrangements in which only one particle contributed to the rearrangement) tended to be bucklers; Figure~3 in the main text suggests that the statistics of these packings are somewhat different from those with higher participation number. Fig.~\ref{figsi:bucklers} shows the rearrangement structure of one such packing.

\bibliography{main}